\documentclass[aps,prl,reprint,superscriptaddress,
		       amsfonts,amssymb,amsmath,showpacs]{revtex4-1}
\usepackage{dcolumn}
\pdfoutput=1
\def\AuthorList{true} 

\usepackage{graphicx}
\usepackage{dcolumn}
\usepackage{amsmath}
\usepackage{bm}
\usepackage{slashed}
\usepackage{rotating}
\usepackage{subfigure}
\usepackage{psfrag}
\usepackage[multidot]{grffile}

\usepackage[tracking=true,kerning=true,expansion=true,spacing=false,factor=1100,stretch=20,shrink=20]{microtype}
\SetTracking{encoding={*}, shape=sc}{50} 

 \usepackage{siunitx}

\usepackage[pdftex,unicode=true,pdfpagelabels=true,pdfusetitle,pdfauthor={The CDF Collaboration}]{hyperref}
\setkeys{Gin}{width=\linewidth,keepaspectratio}

\newcommand{\nnnonw}{NN$_{\mathrm{QCD}}$}
\newcommand{\nnnont}{NN$_{V\mathrm{jets}}$}
\newcommand{\nntt}{NN$_{t\bar{t}}$}
\newcommand{\nnsig}{NN$_{\mathrm{sig}}$}

\begin{document}

\title{\texorpdfstring{\boldmath{Search for $s$-channel Single Top Quark Production in the Missing Energy Plus Jets Sample using the Full CDF II Data Set}}{Search for $s$-channel Single Top Quark Production in the Missing Energy Plus Jets Sample using the Full CDF II Data Set}}

\ifthenelse{\equal{\AuthorList}{true}} { 
\affiliation{Institute of Physics, Academia Sinica, Taipei, Taiwan 11529, Republic of China}
\affiliation{Argonne National Laboratory, Argonne, Illinois 60439, USA}
\affiliation{University of Athens, 157 71 Athens, Greece}
\affiliation{Institut de Fisica d'Altes Energies, ICREA, Universitat Autonoma de Barcelona, E-08193, Bellaterra (Barcelona), Spain}
\affiliation{Baylor University, Waco, Texas 76798, USA}
\affiliation{Istituto Nazionale di Fisica Nucleare Bologna, \ensuremath{^{ii}}University of Bologna, I-40127 Bologna, Italy}
\affiliation{University of California, Davis, Davis, California 95616, USA}
\affiliation{University of California, Los Angeles, Los Angeles, California 90024, USA}
\affiliation{Instituto de Fisica de Cantabria, CSIC-University of Cantabria, 39005 Santander, Spain}
\affiliation{Carnegie Mellon University, Pittsburgh, Pennsylvania 15213, USA}
\affiliation{Enrico Fermi Institute, University of Chicago, Chicago, Illinois 60637, USA}
\affiliation{Comenius University, 842 48 Bratislava, Slovakia; Institute of Experimental Physics, 040 01 Kosice, Slovakia}
\affiliation{Joint Institute for Nuclear Research, RU-141980 Dubna, Russia}
\affiliation{Duke University, Durham, North Carolina 27708, USA}
\affiliation{Fermi National Accelerator Laboratory, Batavia, Illinois 60510, USA}
\affiliation{University of Florida, Gainesville, Florida 32611, USA}
\affiliation{Laboratori Nazionali di Frascati, Istituto Nazionale di Fisica Nucleare, I-00044 Frascati, Italy}
\affiliation{University of Geneva, CH-1211 Geneva 4, Switzerland}
\affiliation{Glasgow University, Glasgow G12 8QQ, United Kingdom}
\affiliation{Harvard University, Cambridge, Massachusetts 02138, USA}
\affiliation{Division of High Energy Physics, Department of Physics, University of Helsinki, FIN-00014, Helsinki, Finland; Helsinki Institute of Physics, FIN-00014, Helsinki, Finland}
\affiliation{University of Illinois, Urbana, Illinois 61801, USA}
\affiliation{The Johns Hopkins University, Baltimore, Maryland 21218, USA}
\affiliation{Institut f\"{u}r Experimentelle Kernphysik, Karlsruhe Institute of Technology, D-76131 Karlsruhe, Germany}
\affiliation{Center for High Energy Physics: Kyungpook National University, Daegu 702-701, Korea; Seoul National University, Seoul 151-742, Korea; Sungkyunkwan University, Suwon 440-746, Korea; Korea Institute of Science and Technology Information, Daejeon 305-806, Korea; Chonnam National University, Gwangju 500-757, Korea; Chonbuk National University, Jeonju 561-756, Korea; Ewha Womans University, Seoul, 120-750, Korea}
\affiliation{Ernest Orlando Lawrence Berkeley National Laboratory, Berkeley, California 94720, USA}
\affiliation{University of Liverpool, Liverpool L69 7ZE, United Kingdom}
\affiliation{University College London, London WC1E 6BT, United Kingdom}
\affiliation{Centro de Investigaciones Energeticas Medioambientales y Tecnologicas, E-28040 Madrid, Spain}
\affiliation{Massachusetts Institute of Technology, Cambridge, Massachusetts 02139, USA}
\affiliation{University of Michigan, Ann Arbor, Michigan 48109, USA}
\affiliation{Michigan State University, East Lansing, Michigan 48824, USA}
\affiliation{Institution for Theoretical and Experimental Physics, ITEP, Moscow 117259, Russia}
\affiliation{University of New Mexico, Albuquerque, New Mexico 87131, USA}
\affiliation{The Ohio State University, Columbus, Ohio 43210, USA}
\affiliation{Okayama University, Okayama 700-8530, Japan}
\affiliation{Osaka City University, Osaka 558-8585, Japan}
\affiliation{University of Oxford, Oxford OX1 3RH, United Kingdom}
\affiliation{Istituto Nazionale di Fisica Nucleare, Sezione di Padova, \ensuremath{^{jj}}University of Padova, I-35131 Padova, Italy}
\affiliation{University of Pennsylvania, Philadelphia, Pennsylvania 19104, USA}
\affiliation{Istituto Nazionale di Fisica Nucleare Pisa, \ensuremath{^{kk}}University of Pisa, \ensuremath{^{ll}}University of Siena, \ensuremath{^{mm}}Scuola Normale Superiore, I-56127 Pisa, Italy, \ensuremath{^{nn}}INFN Pavia, I-27100 Pavia, Italy, \ensuremath{^{oo}}University of Pavia, I-27100 Pavia, Italy}
\affiliation{University of Pittsburgh, Pittsburgh, Pennsylvania 15260, USA}
\affiliation{Purdue University, West Lafayette, Indiana 47907, USA}
\affiliation{University of Rochester, Rochester, New York 14627, USA}
\affiliation{The Rockefeller University, New York, New York 10065, USA}
\affiliation{Istituto Nazionale di Fisica Nucleare, Sezione di Roma 1, \ensuremath{^{pp}}Sapienza Universit\`{a} di Roma, I-00185 Roma, Italy}
\affiliation{Mitchell Institute for Fundamental Physics and Astronomy, Texas A\&M University, College Station, Texas 77843, USA}
\affiliation{Istituto Nazionale di Fisica Nucleare Trieste, \ensuremath{^{qq}}Gruppo Collegato di Udine, \ensuremath{^{rr}}University of Udine, I-33100 Udine, Italy, \ensuremath{^{ss}}University of Trieste, I-34127 Trieste, Italy}
\affiliation{University of Tsukuba, Tsukuba, Ibaraki 305, Japan}
\affiliation{Tufts University, Medford, Massachusetts 02155, USA}
\affiliation{University of Virginia, Charlottesville, Virginia 22906, USA}
\affiliation{Waseda University, Tokyo 169, Japan}
\affiliation{Wayne State University, Detroit, Michigan 48201, USA}
\affiliation{University of Wisconsin, Madison, Wisconsin 53706, USA}
\affiliation{Yale University, New Haven, Connecticut 06520, USA}

\author{T.~Aaltonen}
\affiliation{Division of High Energy Physics, Department of Physics, University of Helsinki, FIN-00014, Helsinki, Finland; Helsinki Institute of Physics, FIN-00014, Helsinki, Finland}
\author{S.~Amerio\ensuremath{^{jj}}}
\affiliation{Istituto Nazionale di Fisica Nucleare, Sezione di Padova, \ensuremath{^{jj}}University of Padova, I-35131 Padova, Italy}
\author{D.~Amidei}
\affiliation{University of Michigan, Ann Arbor, Michigan 48109, USA}
\author{A.~Anastassov\ensuremath{^{v}}}
\affiliation{Fermi National Accelerator Laboratory, Batavia, Illinois 60510, USA}
\author{A.~Annovi}
\affiliation{Laboratori Nazionali di Frascati, Istituto Nazionale di Fisica Nucleare, I-00044 Frascati, Italy}
\author{J.~Antos}
\affiliation{Comenius University, 842 48 Bratislava, Slovakia; Institute of Experimental Physics, 040 01 Kosice, Slovakia}
\author{G.~Apollinari}
\affiliation{Fermi National Accelerator Laboratory, Batavia, Illinois 60510, USA}
\author{J.A.~Appel}
\affiliation{Fermi National Accelerator Laboratory, Batavia, Illinois 60510, USA}
\author{T.~Arisawa}
\affiliation{Waseda University, Tokyo 169, Japan}
\author{A.~Artikov}
\affiliation{Joint Institute for Nuclear Research, RU-141980 Dubna, Russia}
% ...

\collaboration{CDF Collaboration}
\altaffiliation[With visitors from]{
\ensuremath{^{a}}University of British Columbia, Vancouver, BC V6T 1Z1, Canada,
\ensuremath{^{b}}Istituto Nazionale di Fisica Nucleare, Sezione di Cagliari, 09042 Monserrato (Cagliari), Italy,
\ensuremath{^{c}}University of California Irvine, Irvine, CA 92697, USA,
\ensuremath{^{d}}Institute of Physics, Academy of Sciences of the Czech Republic, 182~21, Czech Republic,
\ensuremath{^{e}}CERN, CH-1211 Geneva, Switzerland,
\ensuremath{^{f}}Cornell University, Ithaca, NY 14853, USA,
\ensuremath{^{g}}University of Cyprus, Nicosia CY-1678, Cyprus,
\ensuremath{^{h}}Office of Science, U.S. Department of Energy, Washington, DC 20585, USA,
\ensuremath{^{i}}University College Dublin, Dublin 4, Ireland,
\ensuremath{^{j}}ETH, 8092 Z\"{u}rich, Switzerland,
\ensuremath{^{k}}University of Fukui, Fukui City, Fukui Prefecture, Japan 910-0017,
\ensuremath{^{l}}Universidad Iberoamericana, Lomas de Santa Fe, M\'{e}xico, C.P. 01219, Distrito Federal,
\ensuremath{^{m}}University of Iowa, Iowa City, IA 52242, USA,
\ensuremath{^{n}}Kinki University, Higashi-Osaka City, Japan 577-8502,
\ensuremath{^{o}}Kansas State University, Manhattan, KS 66506, USA,
\ensuremath{^{p}}Brookhaven National Laboratory, Upton, NY 11973, USA,
\ensuremath{^{q}}Queen Mary, University of London, London, E1 4NS, United Kingdom,
\ensuremath{^{r}}University of Melbourne, Victoria 3010, Australia,
\ensuremath{^{s}}Muons, Inc., Batavia, IL 60510, USA,
\ensuremath{^{t}}Nagasaki Institute of Applied Science, Nagasaki 851-0193, Japan,
\ensuremath{^{u}}National Research Nuclear University, Moscow 115409, Russia,
\ensuremath{^{v}}Northwestern University, Evanston, IL 60208, USA,
\ensuremath{^{w}}University of Notre Dame, Notre Dame, IN 46556, USA,
\ensuremath{^{x}}Universidad de Oviedo, E-33007 Oviedo, Spain,
\ensuremath{^{y}}CNRS-IN2P3, Paris, F-75205 France,
\ensuremath{^{z}}Universidad Tecnica Federico Santa Maria, 110v Valparaiso, Chile,
\ensuremath{^{aa}}The University of Jordan, Amman 11942, Jordan,
\ensuremath{^{bb}}Universite catholique de Louvain, 1348 Louvain-La-Neuve, Belgium,
\ensuremath{^{cc}}University of Z\"{u}rich, 8006 Z\"{u}rich, Switzerland,
\ensuremath{^{dd}}Massachusetts General Hospital, Boston, MA 02114 USA,
\ensuremath{^{ee}}Harvard Medical School, Boston, MA 02114 USA,
\ensuremath{^{ff}}Hampton University, Hampton, VA 23668, USA,
\ensuremath{^{gg}}Los Alamos National Laboratory, Los Alamos, NM 87544, USA,
\ensuremath{^{hh}}Universit\`{a} degli Studi di Napoli Federico I, I-80138 Napoli, Italy
}
\noaffiliation
{ \author{The CDF Collaboration} \affiliation{URL http://www-cdf.fnal.gov} }

\date{\today}

\begin{abstract}
 
The first search for single top quark production from the exchange of an $s$-channel virtual {\it W} boson
using events with an imbalance in the total transverse momentum, $b$-tagged jets, and no identified leptons is presented. The full data set collected by the Collider Detector at Fermilab, corresponding to an integrated luminosity of 9.45 fb$^{-1}$ from Fermilab Tevatron proton-antiproton collisions at a center of mass energy of 1.96 TeV, is used. Assuming the electroweak production of top quarks of mass 172.5 GeV/$c^2$ in the $s$-channel, a cross section of $1.12_{-0.57}^{+0.61}$ (stat+syst) pb, with a significance of 1.9 standard deviations, is measured. This measurement is combined with a previous result obtained from events with an imbalance in total transverse momentum, $b$-tagged jets, and exactly one identified lepton, yielding a cross section of $1.36_{-0.32}^{+0.37}$ (stat+syst) pb, with a significance of 4.2 standard deviations.

\end{abstract}


\maketitle


The top quark was discovered at Fermilab in 1995~\cite{1,2} through top-antitop-quark pair production. This process is mediated by the strong interaction and results in the largest contribution to the top-quark-production cross section in hadron colliders. The top quark can also be produced singly via the electroweak interaction involving the $Wtb$ vertex with a {\it W} boson and a $b$ quark. The study of single top quark production is particularly interesting because of the direct dependence of the cross section on the magnitude of the $Wtb$ coupling. Furthermore, electroweak single top quark production from the exchange of an $s$-channel virtual {\it W} boson is of special interest since possible deviations from the standard model (SM) expectation could indicate evidence for non-SM particles such as higher-mass partners of the {\it W} boson ($W'$) or charged Higgs bosons~\cite{8}.
\begin{figure}[h!]
\begin{center}
\includegraphics[width=\linewidth]{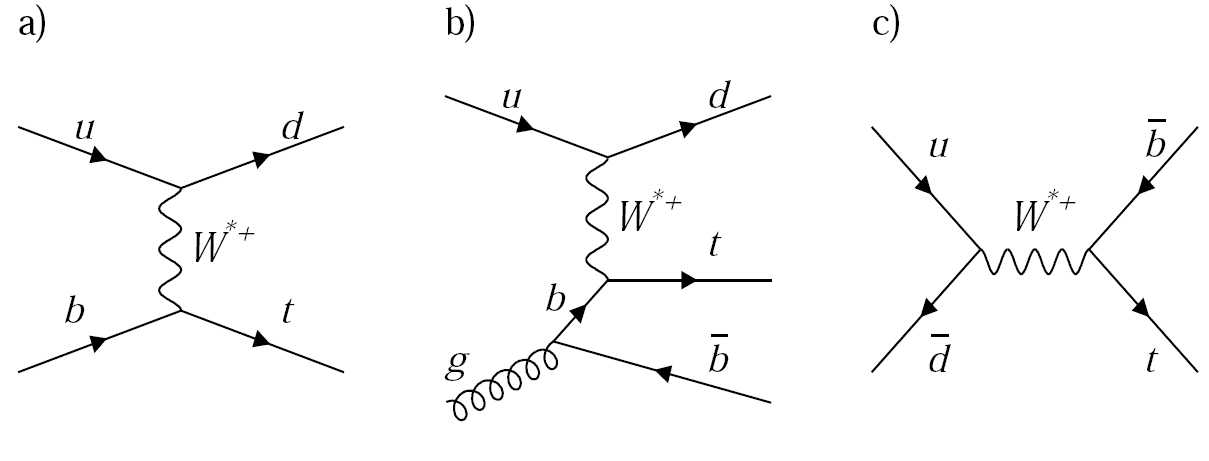}.
\end{center}
\caption{Feynman diagrams for electroweak single top quark production: a) leading-order $t$-channel, b) next-to-leading-order $t$-channel, and c) leading-order $s$-channel.}\label{Feynman}
\end{figure}
Examples of SM single top quark production processes dominating at the Tevatron are shown in Figure~\ref{Feynman}.\\
\indent Single top quark production was observed at the Tevatron in 2009~\cite{3,4,5} in the combined $t$- and $s$-channels. However, $s$-channel production has yet to be observed independently. While the single top quark production through the $t$-channel exchange of a {\it W} boson, first observed by the D0 experiment~\cite{tchannel}, was established in Large Hadron Collider (LHC) proton-proton collisions~\cite{6,7}, the $s$-channel process has an unfavorable production rate compared to the background rates at the LHC. The D0 Collaboration reported the first evidence of $s$-channel single top quark production~\cite{D0}, measuring a cross section of $1.10^{+0.33}_{-0.31}$ (stat+syst) pb, with a significance of 3.7 standard deviations. More recently CDF also obtained 3.8 standard deviation evidence using events containing one isolated muon or electron, large missing transverse energy ($\slashed E_T$)~\cite{coordinates}, and two jets, at least one of which is identified as likely to have originated from a bottom quark ($b$-tagged)~\cite{Hao}. This sample is referred to as the “$\ell \nu b\bar{b}$” sample. In this Letter, a similar search is reported, performed for the first time using an independent sample of events with large $\slashed E_T$, two or three jets of which one or more jets are $b$-tagged, and no detected electron or muon candidates. This sample is referred to as the “$\slashed E_T b\bar{b}$” sample. Most of the techniques developed for the low-mass Higgs boson search~\cite{higgsmetbb} in the same data sample are exploited, including the \textsc{HOBIT} $b$-tagger~\cite{hobit}. By combining the results of the two searches, the best possible sensitivity to $s$-channel single top production from the 9.45 fb$^{-1}$ of integrated luminosity from the full CDF II data set is obtained.\\
\indent In the $\slashed E_T b\bar{b}$ analysis, events are accepted by the online event selection (trigger) that requires $\slashed E_T$ $>$ 45 GeV or, alternatively, $\slashed E_T$ $>$ 35 GeV and two or more jets with transverse energy $E_T$ $>$ 15 GeV. Offline, events containing identified electrons or muons are excluded and $\slashed E_T$ $>$ 35 GeV is required, after correcting measured jet energies for instrumental effects~\cite{Bhatti_2005ai}. Events with two or three high-$E_T$ jets are selected and the two jets with the largest transverse energies, $E_T^{j_1}$ and $E_T^{j_2}$, are required to satisfy $25$ $<$  $E_T^{j_1}$ $<$ 200 GeV and $20$ $<$ $E_T^{j_2}$ $<$ 120 GeV, where the jet energies are determined from calorimeter deposits corrected for track momentum measurements~\cite{h1}. A fraction of events consists of single top quark candidates in which the tau lepton from the $t\rightarrow Wb\rightarrow \tau\nu b$ decay is reconstructed as a jet in the calorimeters. To increase the acceptance for events with an unidentified $\tau$ lepton, events in which the third-most energetic jet satisfies $15\ <\ E_T^{j_3}\ <\ 100$ GeV are accepted. Because of the large rate of inclusive quantum chromodynamics (QCD) multijet (MJ) production, events with four or more reconstructed jets, where each jet has transverse energy in excess of 15 GeV and pseudorapidity~\cite{coordinates} $|\eta|\ <\ 2.4$, are rejected. To ensure that the two leading-$E_T$ jets are within the silicon-detector acceptance, they are required to satisfy $|\eta|\ <\ 2$, with at least one of them satisfying $|\eta| < 0.9$.\\
\indent The MJ background events most often contain $\slashed E_T$ generated through jet energy mismeasurements. Neutrinos produced in semileptonic $b$-hadron decays can also contribute to the $\slashed E_T$ of these events. In both cases, the $\slashed{\vec{E}}_T$ is typically aligned with $\vec{E}_\mathrm{T}^{j_2}$, and events are rejected by requiring the azimuthal separation between $\slashed{\vec{E}}_T$ and $\vec{E}_\mathrm{T}^{j_{2,3}}$ to be larger than 0.4. The remaining MJ background has a large contribution of events with jets from fragmenting light-flavored $u$, $d$, and $s$ quarks or gluons, which can be further reduced by requiring $b$-tagged jets. Charm quarks, which share some features associated with $b$ quarks, are not explicitly identified. Events are assigned to three independent subsamples depending on the \textsc{HOBIT} output of the two leading jets. Jets with \textsc{HOBIT} values larger than 0.98 are defined as tightly tagged (“T-jet”), whereas jets with outputs between  0.72 and 0.98 are defined as loosely tagged (“L-jet”). TT events are defined as those in which both jets are tightly tagged, TL events as those in which one jet is
tightly tagged and the other is loosely tagged, and 1T events as those in which only one jet is tightly tagged while the other is untagged. Events with either two or three jets are analyzed separately, leading to six event subsamples with differing signal to background ratios. This strategy enhances sensitivity and helps separate $s$-channel single top quark production, enhanced in the double-tag categories, from the $t$-channel production, enhanced in the single-tag categories.\\
\indent In order to extract the $s$-channel electroweak single top quark signal from the more dominant background contributions, the rates and kinematic distributions of events associated with each process need to be accurately modeled. The kinematic distributions of events associated with top-quark pair, single top quark, {\it V}+jets (where {\it V} stands for {\it W} and {\it Z} bosons), {\it W+c}, diboson ({\it VV}) and associated Higgs and {\it W} or {\it Z} boson ({\it VH}) production are modeled using simulations. The {\sc alpgen} generator~\cite{alpgen} is used to model {\it V}+jets, {\it W+c}, and {\it VH} production. The {\sc powheg}~\cite{powheg} generator is used to model $t$-and $s$-channel single top quark production, while {\sc pythia}~\cite{pythia} is used to model top-quark-pair and {\it VV} production. Parton showering is simulated in all cases using {\sc pythia}. Event modeling includes simulation of the detector response using {\sc geant}~\cite{geant}. The simulated events are reconstructed and analyzed in the same way as the experimental data. Normalizations of the event contributions from $t$-channel single top quark, {\it VV}, {\it VH}, and $t\bar{t}$ pair production are taken from theoretical cross section predictions~\cite{kido,diboson,hbb_prod,top}, while normalization for {\it W+c} production is taken from the measured cross section~\cite{wc}.  For {\it V}+jets production, the heavy-flavor contribution is normalized based on the number of $b$-tagged events observed in an independent data control sample. Contributions of {\it V}+jets, and {\it VV} events containing at least one incorrectly $b$-tagged, light-flavored jet are determined by applying to simulated events per-event mistag probabilities obtained from a generic event sample containing light-flavored jets~\cite{mistag}. The MJ background~\cite{higgsmetbb} remaining after the full selection criteria is modeled by applying a tag-rate matrix derived from a MJ-dominated data sample to pre-tagged events that otherwise satisfy the signal sample selection criteria. The estimated event yields are shown in Table~\ref{sig2}, \ref{sig3}.\\
\indent In order to separate the $s$-channel single top quark signal from the backgrounds, a staged multivariate neural network (NN) technique is employed. A first network, \nnnonw, is trained to discriminate MJ events from signal events. Events that satisfy a minimal requirement on the \nnnonw\ output variable are further analyzed by a function, \nnsig, derived from the outputs of two additional NNs, \nnnont\ and \nntt, designed respectively to separate the signal from {\it V}+jets (and the remaining MJ events) and $t\bar{t}$ backgrounds.\\
\indent The \nnnonw\ discriminant is trained using QCD multijet events for the background sample and {\it W}+jets events for the signal sample, since the kinematic properties associated with the presence of a {\it W} boson in the $s$-channel single top quark and {\it W}+jets production processes are very similar, in contrast with those of events originating from MJ production. The discriminant is trained separately for the two-jet and three-jet samples using kinematic, angular, and event-shape quantities for the input variables.\\
\indent The two additional networks, \nnnont\ and \nntt, are trained for events that satisfy the minimum requirement on the \nnnonw\ output variable. The first, \nnnont, is trained to separate the $s$-channel single top quark signal from {\it V}+jets and the remaining MJ backgrounds using simulated signal and background made of pretagged data events that satisfy the requirement on \nnnonw, reweighted by the tag-rate parametrization, for the background sample. The \nnnonw\ requirement changes the pretag data composition, enhancing the {\it V}+jets contribution and selecting MJ events with properties closer to those expected for {\it V}+jets events. The background model obtained by reweighting these events via the tag-rate probability accounts for both the {\it V}+jets and MJ event contributions, allowing for more straightforward training of the \nnnont. The second, \nntt, is trained to separate $s$-channel single top quark from $t\bar{t}$ production using simulation for both components. The final discriminant, the \nnsig, is defined as the quadrature-weighted sum of the \nnnont\ and \nntt\ output variables. Figure~\ref{fig:nnsig} shows the predicted and observed shapes of the \nnsig\ output variable for each of the six event subsamples.

\begin{table}[htbp]
  \begin{center}
    \newcolumntype{d}{D{.}{.}{1}}
\caption{\label{sig2}Number of predicted and observed two-jet events in the 1T, TL, and TT subsamples. The uncertainties on the predicted numbers of events are due to the theoretical-cross-section uncertainties and the uncertainties on signal and background modeling. Both the uncertainties and the central values are those obtained from the fit to the data which incorporates the theoretical constraints.}
  \begin{tabular}{c*{3}{c}}
    	\hline
      \hline
      Category                 & 1T & TL & TT \\ 
      \hline           
        $t$-ch single top &    161 $\pm$   31 &     10.8 $\pm$    2.1 &      9.2 $\pm$    1.7\\ 
     $t\bar{t}$ &    243 $\pm$   24 &     84.8 $\pm$    9.3 &     92.4 $\pm$    8.4\\ 
        Diboson &    285 $\pm$    2 &     51.3 $\pm$    0.6 &     37.2 $\pm$    0.5\\ 
           {\it VH} &     12.6 $\pm$    1.4 &      6.6 $\pm$    0.8 &      7.6 $\pm$    0.8\\ 
       {\it V}+jets &   6528 $\pm$ 2048 &    694 $\pm$  216 &    220 $\pm$   69\\ 
             MJ &   8322 $\pm$  180 &    928 $\pm$   59 &    300 $\pm$   32\\
      \hline 
       Signal &     86.2 $\pm$   47.7 &     41.8 $\pm$   23.2 &     45.9 $\pm$   25.3\\
      \hline 
      Total prediction    &     15557 $\pm$ 2056 &   1733 $\pm$  224 &    663 $\pm$   76\\
      \hline 
      Observed    &                   15312  &                 1743  &                  686 \\    
      \hline
      \hline
    \end{tabular}        
  \end{center}
\end{table}

\begin{table}[htbp]
  \begin{center}
    \newcolumntype{d}{D{.}{.}{1}}
    \caption{\label{sig3}Number of predicted and observed three-jet events in the 1T, TL, and TT subsamples. The uncertainties on the predicted numbers of events are due to the theoretical-cross-section uncertainties and the uncertainties on signal and background modeling. Both the uncertainties and the central values are those obtained from the fit to the data which incorporates the theoretical constraints.}
  \begin{tabular}{c*{3}{c}}
    	\hline
      \hline
      Category                 & 1T & TL & TT \\ 
      \hline           
    $t$-ch single top &     82.2 $\pm$   15.8 &      7.5 $\pm$    1.5 &      6.8 $\pm$    1.3\\ 
     $t\bar{t}$ &    597 $\pm$   60 &    118 $\pm$   13 &    110 $\pm$    10\\ 
        Diboson &    108 $\pm$    2 &     15.7 $\pm$    0.4 &      8.8 $\pm$    0.3\\ 
           {\it VH} &      6.0 $\pm$    0.7 &      1.9 $\pm$    0.2 &      2.2 $\pm$    0.2\\ 
       {\it V}+jets &   1610 $\pm$  505 &    165 $\pm$   51 &     50 $\pm$   16\\ 
             MJ &   1818 $\pm$   49 &    188 $\pm$   15 &     55.9 $\pm$    7.6\\
      \hline 
      Signal &     45.7 $\pm$   25.3 &     15.4 $\pm$    8.5 &     16.2 $\pm$    8.9\\ 
      \hline 
      Total prediction   &      4220 $\pm$  511 &    495 $\pm$   55 &    234 $\pm$   20\\ 
      \hline 
      Observed    &              4198  &                  490  &                  237 \\   
      \hline
      \hline
    \end{tabular}        
  \end{center}
\end{table}

\begin{figure}[htbp]
  \begin{minipage}[b]{0.48\linewidth}
    \centering
    \includegraphics[width=\linewidth]{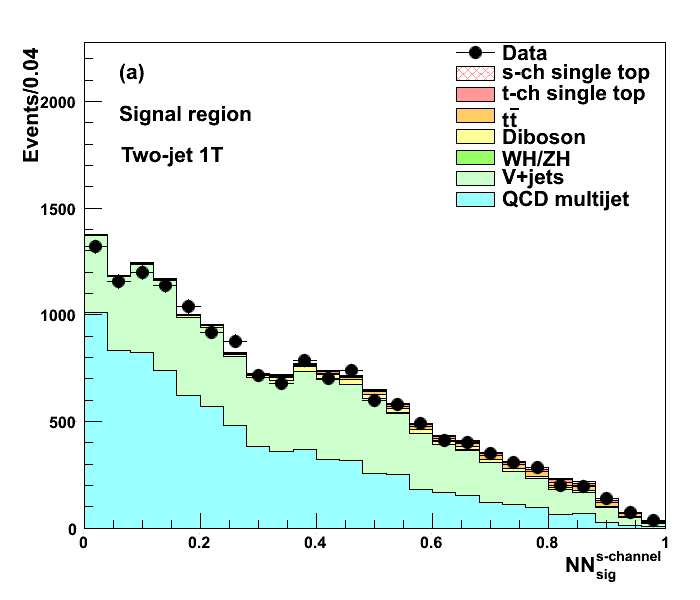}

  \end{minipage}
  \hspace{0.1cm}
  \begin{minipage}[b]{0.48\linewidth}
    \centering
    \includegraphics[width=\linewidth]{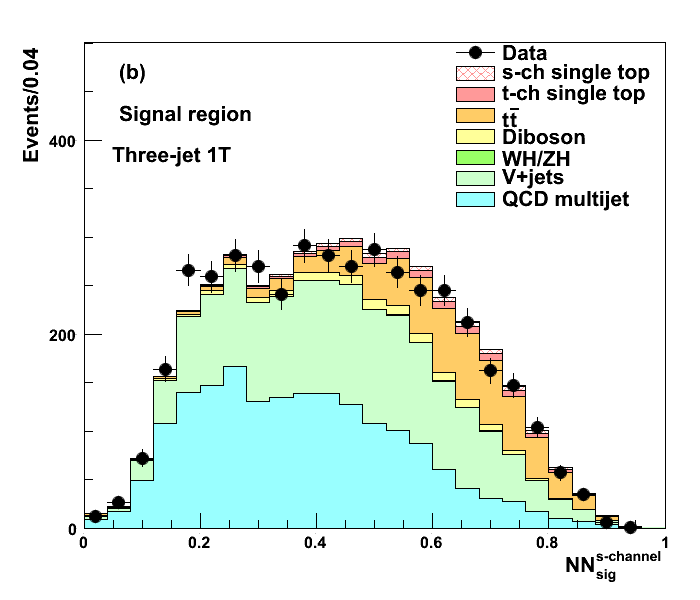}
  
  \end{minipage}\\
  \begin{minipage}[b]{0.48\linewidth}
     \centering
    \includegraphics[width=\linewidth]{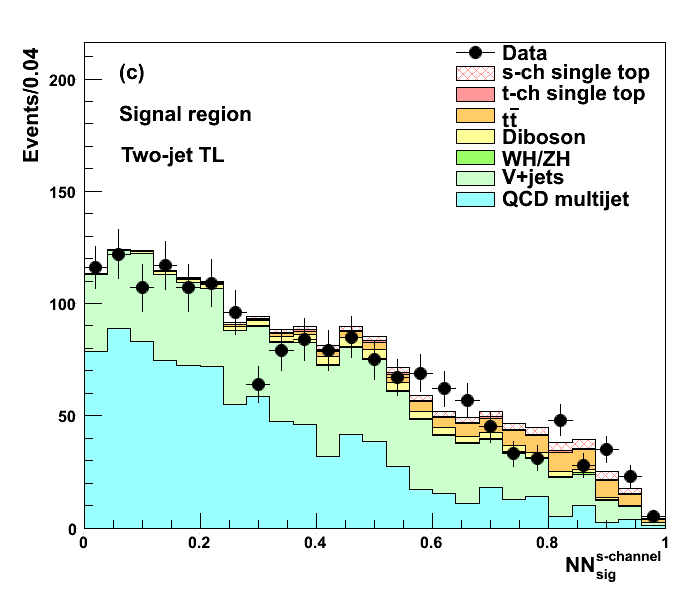}
  
  \end{minipage}
\hspace{0.1cm}
  \begin{minipage}[b]{0.48\linewidth}
    \centering
    \includegraphics[width=\linewidth]{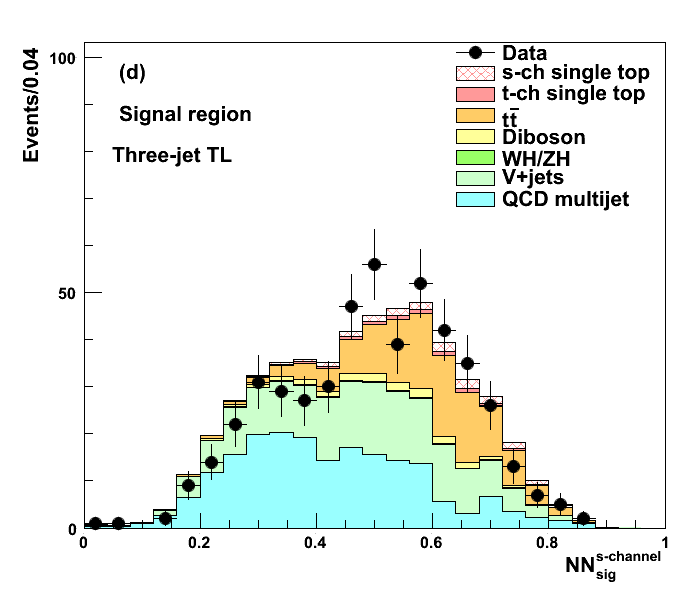}

  \end{minipage}\\
  \begin{minipage}[b]{0.48\linewidth}
    \centering
    \includegraphics[width=\linewidth]{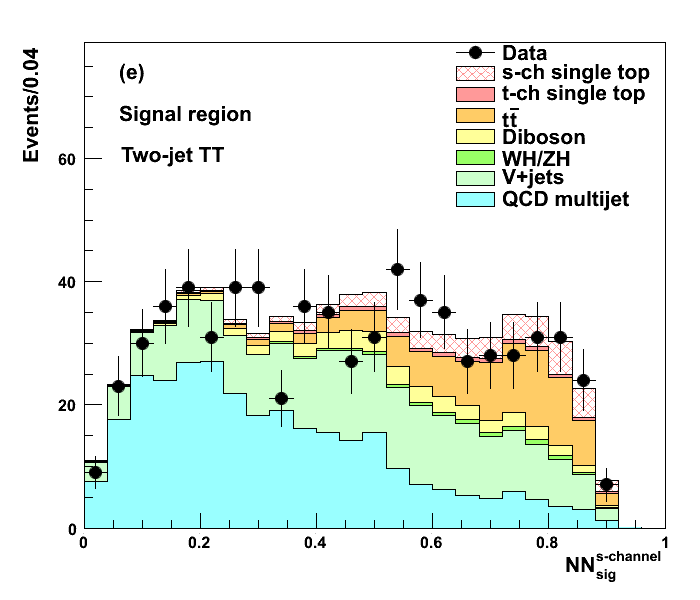}
  
  \end{minipage}
\hspace{0.1cm}
  \begin{minipage}[b]{0.48\linewidth}
    \centering
    \includegraphics[width=\linewidth]{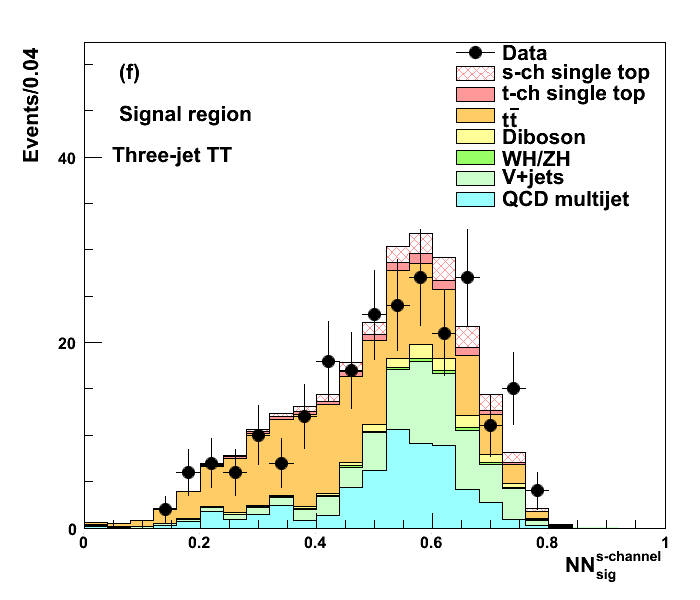}
  
  \end{minipage}
\caption{Predicted and observed final discriminant distributions in the signal region, for (a) 1T two-jet, (b) 1T three-jet, (c) TL two-jet, (d) TL three-jet, (e) TT two-jet and (f) TT three-jet event subsamples.}\label{fig:nnsig}
\end{figure}

The modeling of SM backgrounds is tested in several control samples. A first (EWK) control sample is defined containing events with at least one charged lepton that otherwise satisfy the selection criteria. This sample is independent from the 
signal sample and is sensitive primarily to top-quark pair, {\it V}+jets, and, to a lesser extent, {\it VV} production.
A second (QCD) control sample contains events that do not meet the minimal requirement on the \nnnonw\ output variable but otherwise satisfy the selection criteria.  This event sample, dominated by MJ production, is used to validate the data-driven MJ background model and obtain scale factors, ranging from 0.7 to 0.9, for normalizing modeled contributions to the TT, TL, and 1T event subsamples.  Comparisons of modeled and observed distributions for multiple kinematic variables, including those used as inputs to the \nnnonw, \nnnont, and \nntt, are used to validate the accuracy of the model.\\
\indent To measure the signal contribution, the sum of modeled contributions is fitted to the observed data as a function of the final discriminant variable, \nnsig, accounting for statistical and systematic uncertainties. 
The dominant systematic uncertainties arise from the normalization of the {\it V}-plus-heavy-flavor background
contributions (30\%), differences in $b$-tagging efficiencies between data and simulation (8--16\%), and mistag rates (20--30\%)~\cite{hobit}. Other uncertainties are on the $t\bar{t}$ (3.5\%), $t$-channel single top quark (6.2\%), {\it VV} (6\%), {\it VH} (5\%), and {\it W+c} (23\%) cross sections~\cite{top,kido,diboson,hbb_prod,wc}, normalizations of the QCD multijet background (3--7\%), luminosity measurement (6\%)~\cite{lumi_cdf}, jet-energy scale (1--6\%)~\cite{Bhatti_2005ai}, trigger efficiency (1--3\%), parton distribution functions (2\%), and lepton vetoes (2\%). The shapes obtained by varying the tag-rate probabilities by one standard deviation from their central values are applied as uncertainties on the shapes of the \nnsig output distribution for the MJ background. Changes in the shape of the \nnsig\ distribution originating from jet energy scale uncertainties are also incorporated for processes modeled via the simulation.\\
\indent A likelihood fit to the binned \nnsig\ distribution is used to extract the $s$-channel single top quark signal in the presence of SM backgrounds. The likelihood is the product of Poisson probabilities over the bins of the final discriminant distribution. The mean number of expected events in each bin includes contributions from each background source and $s$-channel single top quark production, assuming a top-quark mass of 172.5 GeV/$c^2$. To extract the signal cross section, a Bayesian method is employed~\cite{limits}. A uniform prior probability in the non-negative range for the $s$-channel single top quark production cross section times branching fraction and truncated Gaussian priors for the uncertainties on the acceptance and shape of each process are incorporated in the fit. Results from each of the six search subsamples are combined by taking the product of their likelihoods and simultaneously varying the correlated uncertainties.\\
\indent The measured $s$-channel single top quark cross section in the $\slashed E_T b\bar{b}$ sample is $1.12_{-0.57}^{+0.61}$ (stat+syst) pb. The probability of observing a signal as large as the observed one or larger that results from fluctuations of the background (p-value) is determined using pseudoexperiments to be is $3.1\ \times\ 10^{-2}$, corresponding to a significance of 1.9 standard deviations. The median expected significance assuming that a signal is present at the SM rate is 1.8 standard deviations.\\
\indent This result is combined with the result of a similar search in the $\ell \nu b\bar{b}$ sample~\cite{Hao}. In that search, candidate events were selected by requiring exactly one reconstructed muon or electron in the final state. Hence, no such events are included in the $\slashed E_T b\bar{b}$ analysis described above. Four independent tagging categories, according to the score of the \textsc{HOBIT} tagger on the two leading jets (tight-tight TT, tight-loose TL, single-tight 1T, and loose-loose LL), were analyzed separately. Events were also divided into three independent categories based on different lepton reconstruction algorithms. 
To further discriminate the signal from all other backgrounds, neural networks were employed. These NNs were optimized separately for each tagging and lepton category. Correlated systematic uncertainties were treated as described above for the $\slashed E_T b\bar{b}$ search. Finally, a Bayesian binned-likelihood technique was applied to the final NN output to extract the $s$-channel single top quark cross section. The significance of the result from the $\ell \nu  b\bar{b}$ channel was 3.8 standard deviations, and the measured cross section was $1.41^{+0.44}_{-0.42}$~(stat+syst) pb, assuming a top-quark mass of 172.5 GeV/$c^2$.\\
\indent The two analyses are combined by taking the product of their likelihoods and simultaneously varying the correlated uncertainties, following the same procedure explained above. The uncertainties associated with the theoretical cross sections of the $t\bar{t}$, $t$-channel electroweak single top quark, {\it VV}, and {\it VH} production processes; the luminosity; the $b$-tagging efficiency; and the mistag rate are considered fully correlated between the two searches. The combined measurement results in an $s$-channel single top quark production cross section of $1.36_{-0.32}^{+0.37}$ pb, consistent with the SM cross section of $1.05 \pm 0.05$ pb~\cite{kido}. The combined p-value is $1.6\ \times\ 10^{-5}$, which corresponds to a signal significance of 4.2 standard deviations. The median expected significance is 3.4 standard deviations.\\
\indent In summary, we perform for the first time a search for $s$-channel single 
top quark production in the $\slashed E_T b\bar{b}$ channel. The result is combined with that 
of a previous search in the $\ell \nu b\bar{b}$ channel~\cite{Hao} to strengthen the 
reported evidence for $s$-channel single top quark production, leading to 
an improvement of more than 10\% on the uncertainty of the measured 
cross section.

\begin{acknowledgments}

We thank the Fermilab staff and the technical staffs of the
participating institutions for their vital contributions. This work
was supported by the U.S. Department of Energy and National Science
Foundation; the Italian Istituto Nazionale di Fisica Nucleare; the
Ministry of Education, Culture, Sports, Science and Technology of
Japan; the Natural Sciences and Engineering Research Council of
Canada; the National Science Council of the Republic of China; the
Swiss National Science Foundation; the A.P. Sloan Foundation; the
Bundesministerium f\"ur Bildung und Forschung, Germany; the Korean
World Class University Program, the National Research Foundation of
Korea; the Science and Technology Facilities Council and the Royal
Society, United Kingdom; the Russian Foundation for Basic Research;
the Ministerio de Ciencia e Innovaci\'{o}n, and Programa
Consolider-Ingenio 2010, Spain; the Slovak R\&D Agency; the Academy
of Finland; the Australian Research Council (ARC); and the EU community
Marie Curie Fellowship Contract No. 302103.

\end{acknowledgments}
\clearpage

\bibliographystyle{apsrev4-1-JHEPfix}

\begin{thebibliography}{99}

\bibitem{1}
F. ~Abe {\it et al.} (CDF Collaboration), Phys. Rev. Lett. {\bf 74}, 2626 (1995).

\bibitem{2}
S. ~Abachi {\it et al.} (D0 Collaboration), Phys. Rev. Lett. {\bf 74}, 2632 (1995).

\bibitem{8}
T. M. P. Tait and C. P. Yuan, Phys. Rev. D {\bf 63}, 014018 (2000).

\bibitem{3}
CDF and D0 Collaboration, Tevatron Electroweak Working Group, arXiv:0908.2171.

\bibitem{4}
T. ~Aaltonen {\it et al.} (CDF Collaboration), Phys. Rev. Lett. {\bf 103}, 092002 (2009).

\bibitem{5}
V. ~M. ~Abazov {\it et al.} (D0 Collaboration), Phys. Rev. Lett. {\bf 103}, 092001 (2009).



\bibitem{tchannel}
V. Abazov {\it et al.} (D0 Collaboration), Phys. Lett. B {\bf 705}, 313 (2011).
\bibitem{6}
G. Aad {\it et al.} (ATLAS Collaboration), Phys. Lett. B {\bf 717}, 330 (2012).
\bibitem{7}
S. Chatrchyan {\it et al.} (CMS Collaboration), J.~High Energy Phys. 12 (2012) 035.
.

%


\bibitem{D0}  V. ~M. ~Abazov {\it et al.} (D0 Collaboration), Phys. Lett. B {\bf 726}, 656 (2013).

\bibitem{coordinates} CDF uses a cylindrical coordinate system with the $z$ axis along the proton beam axis. The pseudorapidity is \mbox{$\eta=-\ln(\tan\frac{\theta}{2})$}, where $\theta$ is the polar angle, and $\varphi$ is the azimuthal angle, 
while \mbox{$p_{\mathrm{T}}=p \sin\theta$} and \mbox{$E_{\mathrm{T}}=E \sin\theta$}. The $\slashed E_T$ is defined as the magnitude of $\slashed{\vec{E}}_T = -\sum_{i} E_{T}^i\hat{n}_i$, where $\hat{n}_i$ is a unit vector perpendicular to the beam axis and pointing at the $i$th calorimeter tower, and $E_{T}^i$ is the transverse energy therein.

\bibitem{Hao}
T. ~Aaltonen {\it et al.} (CDF Collaboration), arXiv:1402.0484.



\bibitem{higgsmetbb} 
  T.~Aaltonen {\it et al.}  (CDF Collaboration), Phys.\ Rev.\ D {\bf 87}, 052008 (2013).

\bibitem[{hob()}]{hobit}

\bibinfo{note}{J.~Freeman, T.~Junk, M.~Kirby, Y.~Oksuzian, T.~J.~Phillips,
  F.~D.~Snider, M.~Trovato, J.~Vizan, and W.~M.~Yao, Nucl. Instrum. Methods
  Phys. Res., Sect. A {\bf 697}, 64 (2013).}

\bibitem[{Bha()}]{Bhatti_2005ai}
\bibinfo{note}{A.~Bhatti {\it et al.}, Nucl. Instrum. Methods Phys. Res., Sect.
  A {\bf 566}, 375 (2006).}

\bibitem[{h1()}]{h1}
\bibinfo{note}{C.~Adloff {\it et al.} (H1 Collaboration), Z. Phys. C {\bf 74},
  221 (1997).}

\bibitem{alpgen}M.L.~Mangano, M.~Moretti, F.~Piccinini, R.~Pittau, and A.D.~Polosa, J.~High Energy Phys. 0307 (2003) 001. 
\bibitem{powheg}S.~Alioli, P.~Nason, C.~Oleari, and E.~Re,	J.~High Energy Phys.\ 06 (2010) 043. 
\bibitem{pythia}
T.~Sjostrand, S.~Mrenna, and P.~Skands, J.~High Energy Phys. 05 (2006) 026.

\bibitem{geant} GEANT, detector description and simulation tool, CERN Program Library Long Writeup W5013 (1993).


\bibitem[{top({\natexlab{a}})}]{top}
\bibinfo{note}{P. ~Baernreuther, M. ~Czakon and A. ~Mitov, Phys. Rev. Lett. {\bf 109}, 132001 (2012).}

\bibitem{kido}
N.~Kidonakis, Phys. Rev. D {\bf 81}, 054028 (2010).

\bibitem[{dib()}]{diboson}
\bibinfo{note}{J.~M.~Campbell and R.~K.~Ellis, Phys. Rev. D {\bf 60}, 113006
  (1999)}.


\bibitem[{hbb({\natexlab{b}})}]{hbb_prod}
\bibinfo{note}{J.~Baglio and A.~Djouadi, J. High Energy Phys. 10 (2010) 064;
  O.~Brien, R.~V.~Harlander, M.~Weisemann, and T.~Zirke, Eur. Phys. J. C {\bf
  72}, 1868 (2012).}

\bibitem{wc}
T. ~Aaltonen {\it et al.} (CDF Collaboration), Phys. Rev. Lett. {\bf 110}, 071801 (2013).

\bibitem{mistag}
D.~Acosta {\it et al.} (CDF Collaboration), Phys. Rev. D {\bf 71}, 052003 (2005); A.~Abulencia {\it et al.} (CDF Collaboration), Phys. Rev. D {\bf 74}, 072006 (2006).


%


\bibitem[{lum()}]{lumi_cdf}
\bibinfo{note}{S.~Klimenko, J.~Konigsberg, and T.~M.~Liss, Report No.
  FERMILAB-FN-0741, 2003.}


\bibitem{limits} 
{\it Statistics}, in K. Nakamura {\it et al.} (Particle Data Group), J. Phys. G {\bf 37} 075021 (2010).




\end{thebibliography}

\clearpage


\clearpage

\end{document}